\documentclass[11pt]{article}

\usepackage[dvips]{graphicx}
\usepackage[english]{babel}
\usepackage{psfrag}

\newcommand{\tr}{\mathop{\rm tr}\nolimits}

\newcommand{\II}{\hbox{{1}\kern-.25em\hbox{l}}}

\newcommand{\half}{ {\textstyle\frac{1}{2}} }

\newcommand{\lrD}{{D^{\hspace{-0.8em}
      \raisebox{0.8ex}{$\scriptstyle\leftrightarrow$}}}{}}
\newcommand{\lD}{{D^{\hspace{-0.8em}
      \raisebox{0.8ex}{$\scriptstyle\leftarrow$}}}{}}
\newcommand{\rD}{{D^{\hspace{-0.8em}
      \raisebox{0.8ex}{$\scriptstyle\rightarrow$}}}{}}

\newcommand{\lrpartial}{{\partial^{\hspace{-0.65em}
      \raisebox{0.8ex}{$\scriptstyle\leftrightarrow$}}}{}}
\newcommand{\lpartial}{{\partial^{\hspace{-0.65em}
      \raisebox{0.8ex}{$\scriptstyle\leftarrow$}}}{}}
\newcommand{\rpartial}{{\partial^{\hspace{-0.65em}
      \raisebox{0.8ex}{$\scriptstyle\rightarrow$}}}{}}

\begin{document}
\begin{titlepage}
\begin{flushright}
\begin{tabular}{l}
DESY-05-081 \\
hep-ph/0505269
\end{tabular}
\end{flushright}

\vskip3cm
\begin{center}

\textbf{\LARGE Generalized parton distributions for the pion \\[0.3em]
in chiral perturbation theory}

\vspace*{1.6cm}

{\large
M. Diehl$^1$,
A. Manashov$^{2,3}$
 and
A.~Sch\"afer$^2$
}

\vspace*{0.4cm}

{\sl
$^1$ Deutsches Elektronen-Synchroton DESY, D-22603 Hamburg, Germany \\
$^2$ Institut f\"ur Theoretische Physik, Universit\"at
                          Regensburg, D-93040 Regensburg, Germany \\
$^3$ Department of Theoretical Physics,  Sankt-Petersburg State University,
St.-Petersburg, Russia
}

\vspace*{0.8cm}

{\bf Abstract\\[10pt]} 
\parbox[t]{\textwidth}{ 
Generalized parton distributions provide a unified parameterization of
hadron structure and allow one to combine information from many
different observables.  Lattice QCD calculations already provide
important input to determine these distributions and hold the promise
to become even more important in the future.  To this end, a reliable
extrapolation of lattice calculations to the physical quark and pion
masses is needed.  We present an analysis for the moments of
generalized parton distributions of the pion in one-loop order of
chiral perturbation theory.}

\vspace{1cm}

\end{center}

\end{titlepage}

\section{Introduction}
\setcounter{footnote}{0}

Generalized parton distributions (GPDs) provide general
parameterizations for non-forward hadronic matrix elements, see e.g.\
the original work \cite{GPDs,Ji,R97} and the recent reviews
\cite{GPV,Diehl03,Belitsky:2005qn}.  They contain the usual parton
distribution functions and elastic form factors, and they are closely
related with light-cone wave functions and distribution
amplitudes. GPDs can be analyzed using standard operator product
expansion techniques \cite{GPDs,blumlein}, and factorization theorems
have been proven for many different processes.  In addition, evolution
equations have been derived to next-to-leading order accuracy, and
generally speaking the GPD formalism has reached a similar level of
stringency as the QCD description of inclusive deep inelastic
scattering.  Finally, it has become clear that GPDs contain
information which cannot be deduced directly from any experiment, most
notably the orbital angular momentum carried by partons \cite{Ji} and
the transverse structure of hadrons in the impact parameter plane
\cite{burkardt}.  On the experimental side, data which is especially
suited to constrain the form of GPDs has been obtained at DESY
\cite{DESY} and Jefferson Lab \cite{JLab}.

An important problem hampering the practical realization of the
physics potential of GPDs is that, typically, measurements are only
sensitive to convolutions of GPDs, such that the assumed functional
form enters crucially into the extraction of the distributions
themselves.  The situation is further complicated by potentially large
higher-twist contributions in some processes.  Therefore, the
possibility to obtain independent information from lattice QCD, which
allows one to directly calculate $x$-moments of GPDs, is especially
interesting.  Pioneering lattice calculations for nucleon GPDs have
already been performed some time ago \cite{QCDSF}. These calculations
are being improved and their scope is being systematically widened
\cite{LatticeGPD}, also in the direction of analyzing other hadrons,
in particular the pion.  Calculations for the pion are being performed
by the QCDSF collaboration, using dynamical $N_F=2$ improved Wilson
fermions, but have to be made at unphysically large quark masses to
save on computing time.  The reliable extrapolation to the chiral
limit, i.e.\ to physical quark masses, requires precise knowledge on
the functional form of the quark mass dependence of moments of GPDs.
Close to the chiral limit, this knowledge is provided by chiral
perturbation theory (ChPT), and in recent years the combination of
ChPT and lattice QCD has in fact proven to be very powerful in
obtaining precise lattice predictions.  Obviously the ChPT analysis
should be performed in parallel to the lattice calculations, and this
is the objective of our paper.

In this Letter we present an analysis for the moments of pion GPDs in
one-loop order of ChPT. The paper is organized as follows. In
Section~\ref{sec:definitions} we recall the definition of the pion
GPDs and the expression of their moments in terms of the matrix
elements of local twist-two operators.  In Sections~\ref{sec:tensor}
and \ref{sec:higher-order} we construct the relevant twist-two
operators in the first two orders of ChPT, and in
Section~\ref{sec:one-loop} we calculate the one-loop contributions to
their matrix elements in the pion.  We summarize our findings in
Section~\ref{sec:sum}.

\section{Definitions and properties of pion GPDs}
\label{sec:definitions}

The pion GPDs can be introduced as the pion matrix elements of
nonlocal operators.  As is done in lattice QCD calculations, we assume
isospin symmetry to be exact, neglecting the difference between $u$-
and $d$-quark masses.  We shall study the isoscalar and isovector
quark GPDs, which are respectively defined as
\begin{eqnarray}
  \label{pion-gpd}
\delta^{ab}\, H^{I=0}(x,\xi,t) &=&
\frac{1}{2} \int \frac{d \lambda}{2\pi}\, e^{ix \lambda (P u)}\,
  \langle \pi^b(p')|\, \bar q (-\half\lambda u)\, 
  u^\mu \gamma_\mu\, q(\half\lambda u)\,
  \,|\pi^a(p) \rangle \, ,
\\[2mm]
\label{pion-gpd-isov}
{i\epsilon^{abc}}\, H^{I=1}(x,\xi,t) &=&
\frac{1}{2} \int \frac{d \lambda}{2\pi}\, e^{ix \lambda (P u)}\,
  \langle \pi^b(p')|\,  \bar q (-\half\lambda u)\, 
  u^\mu \gamma_\mu\, \tau^c\, q(\half\lambda u)\,
  \,|\pi^a(p) \rangle \, ,
\end{eqnarray}
where $q$ denotes the doublet of $u$- and $d$-quark fields and
$\tau^a$ the Pauli matrices in isospin space.  $u$~is a lightlike
auxiliary vector, and we use the standard notations for the
kinematical variables
\begin{eqnarray}
P=\frac{p+p'}{2} , \qquad\Delta=p'-p, 
\qquad t=\Delta^2 , \qquad \xi=-\frac{\Delta u}{2Pu} \,.
\end{eqnarray}
In terms of the distributions for individual quark flavors defined in
\cite{Diehl03} we have $H^{I=0} = H^u_\pi + H^d_\pi$ and $H^{I=1} =
H^u_\pi - H^d_\pi$.  As usual, Wilson lines between the quark fields
are to be inserted in (\ref{pion-gpd}) and~(\ref{pion-gpd-isov}) if
one is not working in the light-cone gauge $u^\mu A_\mu =0$.

Time reversal and charge conjugation symmetry result in the following
properties for the pion GPDs:
\begin{eqnarray}
\label{symrel-1}
H^I(x,\xi,t)&=&H^I(x,-\xi,t)\,,\\[1mm]
\label{symrel-2}
H^I(x,\xi,t)&=&(-1)^{1-I} H^I(-x,\xi,t)\,.
\end{eqnarray}
The distributions can be analytically continued to positive $t$ and
are real-valued for $t<4m_\pi^2$, as discussed in
\cite{Diehl:1998dk,Polyakov:1998ze}.

The $x$-moments of the GPDs are related to the matrix elements of the
local isoscalar and isovector operators
\begin{equation}
\label{tensor-ops}
\mathcal{O}^s_{\mu_1 \mu_2 \ldots \mu_n} =
  \mathbf{S}\, \frac{1}{2} \bar{q} \gamma_{\mu_1}
       i \lrD_{\mu_2} \ldots i \lrD_{\mu_n}\, q \, , \qquad
\mathcal{O}^a_{\mu_1 \mu_2 \ldots \mu_n} =
  \mathbf{S}\, \frac{1}{2} \bar{q} \gamma_{\mu_1}
       i \lrD_{\mu_2} \ldots i \lrD_{\mu_n} \tau^a\, q \, ,
\end{equation}
where $\lrD_\mu = \half (\rD_\mu - \lD_\mu) = \half
(\hspace{0.2pt}\rpartial_\mu - \lpartial_\mu) - ig A_\mu$ and all
fields are to be taken at the same space-time coordinate.
$\mathbf{S}$ denotes symmetrization and subtraction of trace terms for
all uncontracted Lorentz indices.  Defining form factors
\begin{eqnarray}
\label{ff-defs}
\langle \pi^b(p')|\, \mathcal{O}^s_{\mu_1 \mu_2 \ldots \mu_n} |
\pi^a(p) \rangle &=& \delta^{ab} \,
\sum_{k=0, 2, \ldots}^{n} A_{n,k}^{I=0}(t)\,
  \mathbf{S}\, \Delta_{\mu_1} \ldots \Delta_{\mu_k} 
               P_{\mu_{k+1}} \ldots P_{\mu_n} \,,
\\
\langle \pi^b(p')|\, \mathcal{O}^c_{\mu_1 \mu_2 \ldots \mu_n} |
\pi^a(p) \rangle &=& i\epsilon^{abc} \,
\sum_{k=0, 2, \ldots}^{n}  A_{n,k}^{I=1}(t)\,
  \mathbf{S}\, \Delta_{\mu_1} \ldots \Delta_{\mu_k} 
               P_{\mu_{k+1}} \ldots P_{\mu_n} \,,
\end{eqnarray}
where the operators are to be taken at zero space-time position and
where time reversal invariance limits $k$ to even values, one finds
\begin{equation}
\label{GPD-exp}
\int_{-1}^{1} dx\, x^{n-1}\, H^{I}(x,\xi,t) =
\sum_{k=0,2,\ldots}^{n} (2\xi)^k\, A^{I}_{n,k}(t) \,.
\end{equation}
Note that because of the symmetry relation (\ref{symrel-2}) the
moments with odd $n$ in (\ref{GPD-exp}) vanish for $I=0$ and those
with even $n$ vanish for $I=1$.  The form factors $A_{n,k}^I(t)$ are
real-valued in the region $t<4m_\pi^2$.  Information on the lowest
moments
\begin{eqnarray}
\label{ff}
\int_{-1}^{1} dx\, H^{I=1}(x,\xi,t)
&=& A_{1,0}^{I=1}(t)\,,\\[1mm]
\int_{-1}^{1} dx\,x\, H^{I=0}(x,\xi,t)
&=& A_{2,0}^{I=0}(t)+ 4\xi^2 A_{2,2}^{I=0}(t)
\end{eqnarray}
will soon be accessible from the lattice for pion masses down to
values $m_\pi\sim 500$~MeV ~\cite{Dirk}.  Note that the first moment
of $H^{I=1}$ is related to the pion electromagnetic form factor as
$A_{1,0}^{I=1}(t)=2F^{\mathit{em}}_\pi(t)$ and the second moment of
$H^{I=0}$ with the form factors in the parameterization \cite{DL91}
\begin{eqnarray}
\label{en-mom}
\langle \pi^b(p')|\, T_{\mu\nu}^Q \,|\pi^a(p) \rangle
= \delta^{ab} \Bigg[
2P_\mu P_\nu\, \theta_2(t)+\frac12
\left( g_{\mu\nu}\Delta^2-\Delta_\mu\Delta_\nu\right) \theta_1(t)
\Bigg] \,,
\end{eqnarray}
where $T^Q_{\mu\nu} = \half \bar{q}\, ( \gamma_\mu\,i\lrD_\nu +
\gamma_\nu\, i\lrD_\mu )\, q$ is the quark part of the energy-momentum
tensor.  Comparison with (\ref{ff-defs}) gives
$\theta_2(t)=A_{2,0}^{I=0}(t)$ and $\theta_1(t)=-4A_{2,2}^{I=0}(t)$.

An analogous discussion can be given for the gluon GPD, which is
defined by
\begin{equation}
  \label{gluon-gpd}
\delta^{ab}\, H^{g}_\pi(x,\xi,t) = 
\frac{1}{Pu} \int \frac{d \lambda}{2\pi}\, e^{ix \lambda (P u)}\,
  \langle \pi^b(p')|\, u_\alpha G^{\alpha\mu}(-\half\lambda u)\, 
  u_\beta G_\mu{}^\beta(\half\lambda u)
  \,|\pi^a(p) \rangle 
\end{equation}
in the convention of \cite{Diehl03}.  It has the same symmetry
properties as $x H^{I=0}(x,\xi,t)$, and its moments $\int dx\, x^{n-2}
H_\pi^g(x,\xi,t)$ are analogs of the moments $\int dx\, x^{n-1}
H^{I=0}(x,\xi,t)$, with which they mix under renormalization scale
evolution.  For ease of notation we will in the following concentrate
on the quark sector, bearing in mind that our results for the
isoscalar form factors $A^{I=0}_{n,k}(t)$ readily generalize to their
gluonic counterparts.

\section{Tensor operators in ChPT}
\label{sec:tensor}

We use the standard ${\cal O}(p^n)$ power-counting of ChPT, where $p$
denotes a generic pion energy or momentum.  The Goldstone boson fields
are collected in the matrix-valued field $U$, and the leading-order
Lagrangian, which is of order ${\cal O}(p^2)$, reads \cite{GL}
\begin{eqnarray}
\label{LO}
{\cal L}_{\pi\pi}^{(2)}&=&\frac{1}{4} F^2
\tr\left(D_\mu U D^\mu U^\dagger+\chi^\dagger U+\chi U^\dagger\right)
\end{eqnarray}
for two light flavors, with the covariant derivative $D_\mu$ and the
external field tensor $\chi$ defined as
\begin{eqnarray}
D_\mu U&=& \partial_\mu U-i{r}_\mu\,U+iU\,{l}_\mu \,, \\
\chi   &=& 2B_0\left({s}+i\,{p}\right) \,.
\end{eqnarray}
Here ${s}$ and ${p}$ denote external scalar and pseudoscalar fields,
which count as quantities of order ${\cal O}(p^2)$.  Furthermore,
${r}_\mu$ and ${l}_\mu$ are external right- and left-handed vector
fields with intrinsic chiral power ${\cal O}(p)$.  The two
leading-order parameters appearing in (\ref{LO}) are the pion decay
constant $F$ (normalized to $F \approx 92$~MeV) and the two-flavor
chiral condensate $B_0=- \langle 0|\bar{q}q|0 \rangle /F^2$, both
evaluated in the chiral limit \cite{GL}.  Throughout this work we use
the non-linear representation for the pion fields $\pi^a$,
\begin{eqnarray}
U=\exp\{i\tau^{a}\pi^{a}/F \} \,.
\end{eqnarray}
Left and right chiral rotations of the nonlinear pion field
\begin{equation}
U \rightarrow V_R^{\phantom{\dagger}}\,U\,V_L^\dagger 
\end{equation}
induce the following transformation of the external fields:
\begin{eqnarray}
r_\mu &\rightarrow & V_R^{\phantom{\dagger}}\,r_\mu\,V_R^\dagger
   +i\,V_R^{\phantom{\dagger}}\,\partial_\mu V_R^\dagger \,,
\qquad\qquad
\chi \:\rightarrow\: V_R^{\phantom{\dagger}}\,\chi\,V_L^\dagger \,,
\\
l_\mu &\rightarrow & V_L^{\phantom{\dagger}}\,l_\mu\,V_L^\dagger
   +i\,V_L^{\phantom{\dagger}}\,\partial_\mu V_L^\dagger \,. 
\nonumber 
\end{eqnarray}

One method to calculate the matrix elements of the operators
$\mathcal{O}^{s}_{\mu_1 \mu_2 \ldots \mu_n}$ and
$\mathcal{O}^{a}_{\mu_1 \mu_2 \ldots \mu_n}$ from (\ref{tensor-ops})
in ChPT would be to introduce corresponding tensor currents as
external sources in the Lagrangian, in generalization of the vector
currents $r_\mu$ and $l_\mu$ in (\ref{LO}).  For tensors of higher
rank, this becomes however increasingly awkward.  Instead, we will
directly match the operators in (\ref{tensor-ops}) onto operators with
pion degrees of freedom.  The expression of the isovector vector
current is for instance well-known to be~\cite{GL}
\begin{eqnarray}\label{isov}
{\cal O}_\mu^a = \frac{1}{2} \bar q \gamma_\mu \tau^a\, q
  &\simeq &
V_\mu^a=-\frac{i}{4} F^2
  \tr \left\{\tau^a\left(U^\dagger \partial_\mu U+
             U\partial_\mu U^\dagger\right)\right\} 
\end{eqnarray}
to leading order in the chiral expansion.  Despite the progress in
deriving chiral dynamics directly from QCD (see e.g.\ \cite{eff}) this
matching problem is far from being solved in general.  The so-called
low-energy constants appearing in ChPT can in principle be determined
by fits to either experimental or lattice data, but only a limited
number of them are known in practice.  What we can however do is to
construct for a given quark or gluon operator in QCD all operators in
ChPT with the same symmetry.  Their matching coefficients typically
are linear combinations of low-energy constants and can be fitted
directly to lattice data at different unphysical pion masses, provided
they are sufficiently close to the chiral limit.  In
Section~\ref{sec:one-loop} we will determine the functional dependence
on $m_\pi^2$ of the form factors $A^{I}_{n,k}(t)$ required for such a
fit at one-loop level in ChPT.

For the discussion of chiral symmetry it is useful to consider the
left- and right-handed tensor operators
\begin{eqnarray}
  \label{OL-def}
({\cal O}^{L}_{ij})_{\mu_1\ldots\mu_n} &=&
   \mathbf{S}\, \bar{q}_j \gamma_{\mu_1} \frac{1-\gamma_5}{2}\,
       i \lrD_{\mu_2} \ldots i \lrD_{\mu_n}\, q_i ,
\\
  \label{OR-def}
({\cal O}^{R}_{ij})_{\mu_1\ldots\mu_n} &=&
   \mathbf{S}\, \bar{q}_j \gamma_{\mu_1} \frac{1+\gamma_5}{2}\,
       i \lrD_{\mu_2} \ldots i \lrD_{\mu_n}\, q_i ,
\end{eqnarray}
which take values in two-dimensional isospin space.  The operators in
(\ref{tensor-ops}) are then given by
\begin{equation}\label{ea}
{\cal O}^s_{\mu_1\ldots\mu_n}= \frac{1}{2} 
 \mathrm{tr}\,\Big\{ {\cal O}^{L}_{\mu_1\ldots\mu_n}+
{\cal O}^{R}_{\mu_1\ldots\mu_n} \Big\}\,, \qquad
{\cal O}^a_{\mu_1\ldots\mu_n}= \frac{1}{2} 
 \mathrm{tr}\,\Big\{\tau^a \left({\cal O}^{L}_{\mu_1\ldots\mu_n}+
{\cal O}^{R}_{\mu_1\ldots\mu_n}\right) \Big\}\,.
\end{equation}
Note that for ease of notation we shall use the same symbol for the
quark operators in QCD and for the operators representing them in
ChPT.  The left- and right-handed operators behave like
\begin{equation}
{\cal O}^{L}_{\mu_1\ldots\mu_n}\to 
  V_{L}^{\phantom{\dagger}}{\cal O}^{L}_{\mu_1\ldots\mu_n}
  V_{L}^\dagger , \qquad\qquad
{\cal O}^{R}_{\mu_1\ldots\mu_n}\to 
  V_{R}^{\phantom{\dagger}}{\cal O}^{R}_{\mu_1\ldots\mu_n}
  V_{R}^\dagger
\end{equation}
under chiral rotations and are related as ${\cal
O}^{R}_{\mu_1\ldots\mu_n}=P{\cal O}^{L}_{\mu_1\ldots\mu_n}P^{-1}$ by
the parity transformation $P$.  To construct a linearly independent
set of operators for the expansion of
$\mathcal{O}^L_{\mu_1\ldots\mu_n}$, let us inspect the building blocks
at hand.  For the purpose of calculating matrix elements between pion
states we can set the external fields $p$, $r_\mu$ and $l_\mu$ to
zero, and keep the scalar field $s$ as a nonzero constant field
implementing the explicit breaking of chiral symmetry by the quark
masses.  Since we assume exact isospin symmetry, this simply gives
$\chi=m^2\, \II$, where $m$ is the bare pion mass in ChPT.  For
constructing the operator this leaves us with the objects $U$, $\chi$
and $\partial_\mu$, with the additional condition
$\partial_\mu\chi=0$.  It is convenient to introduce the following
quantities
\begin{equation}\label{bb}
L_\mu=U^\dagger \partial_\mu U , \qquad\qquad
\chi^L_\pm=\chi^\dagger U\pm U^\dagger \chi \,,
\end{equation}
which transform covariantly under left rotations and are invariant
under right rotations.  The product of some number of $L_\mu$ and
$\chi^L_\pm$ with any number of additional derivatives acting on them
transforms covariantly under left rotations.  Since $\partial_\mu
\chi^L_\pm= \half [\chi_\pm^L,L_\mu]+ \half \{\chi_\mp^L,L_\mu\}$ for
$\partial_\mu\chi=0$, one can eliminate all derivatives acting on
$\chi_\pm^L$.  Similarly, taking into account the identities
\begin{equation}\label{fmn}
\partial_\mu L_\nu-\partial_\nu L_\mu=[L_\mu,L_\nu] , \qquad\qquad
\partial_\mu \partial^\mu L_\nu 
= \partial_\nu \partial^\mu L_\mu
  +[\partial^\mu L_\nu, L_\mu]+[L_\nu, \partial^\mu L_\mu]
\end{equation}
and the equation of motion for the field $U$,
\begin{equation}\label{EQM}
\partial^\mu L_\mu = -\frac{1}{4} \Big( 2\chi_{-}^L
                     - \II \tr \chi_{-}^L\Big) \,,
\end{equation}
one can trade the tensors $\partial_{\nu_1}\ldots\partial_{\nu_{n-1}}
L_{\nu_n}$ for the symmetric and traceless ones,
\begin{equation}
  \label{L-tensor-def}
L_{\nu_1\ldots\nu_n} = \mathbf{S}\,
  \partial_{\nu_1}\ldots\partial_{\nu_{n-1}} L_{\nu_n} .
\end{equation}
A linearly independent set of operators transforming like ${\cal
O}^{L}_{\mu_1\ldots\mu_n}$ under chiral rotations can hence be
constructed from products of the fields $\chi_\pm^L$ and the tensors
$L_{\nu_1\ldots\nu_k}$.  In the ChPT expansion of ${\cal
O}^{L}_{\mu_1\ldots\mu_n}$ they can be multiplied by chirally
invariant operators, which can in turn be constructed from the traces
of products of $\chi_\pm^L$ and
$L_{\nu_1\ldots\nu_k}$.\footnote{Equivalently, one may take the traces
of products of $\chi_\pm^R$ and $R_{\nu_1\ldots\nu_k}$ introduced
below.}
In each term of the expansion, some Lorentz indices of the tensors
$L_{\nu_1\ldots\nu_k}$ are equal to $\mu_1,\ldots,\mu_n$ and all
others are contracted.  Since $\partial_\mu$ and $L_\mu$ are of order
$\mathcal{O}(p)$, each contraction of the indices, as well as the
insertion of a field $\chi_\pm^L$ increases the chiral counting by a
factor ${\cal O}(p^2)$.

At leading order of ChPT the operator ${\cal
O}^{L}_{\mu_1\ldots\mu_n}$ can hence be represented as the sum of
products of $L$ tensors and their traces, with no contracted Lorentz
indices.  It is convenient to contract free indices with the
light-line vector $u$, introducing the notation
\begin{equation}
\label{u-notation}
{\mathcal O}_n^L(u)
  ={\mathcal O}^L_{\mu_1\ldots \mu_n}u^{\mu_1}\ldots u^{\mu_n} , 
\qquad\qquad
L_k(u)=L_{\mu_1\ldots\mu_k}u^{\mu_1}\ldots u^{\mu_k} .
\end{equation}
The operator ${\mathcal O}_n^L(u)$ is then expanded on terms of the
form
\begin{equation}\label{gt}
L_{k_1}(u) \ldots L_{k_{j_1}}(u)\, 
\tr\Big\{L_{k_{j_{1}+1}}(u) \ldots L_{k_{j_2}}(u)\Big\} \,
\tr\Big\{L_{k_{j_{2}+1}}(u) \ldots L_{k_{j_3}}(u)\Big\} \ldots 
\end{equation}
and explicitly reads
\begin{equation}\label{OL}
{\cal O}^{L}_n(u) = c_{n}\, L_n(u)+ \sum_{k=1}^{n-1}
c_{n,k}\,L_k(u)\,L_{n-k}(u) + \ldots \,,
\end{equation}
where we used that $\tr L_\mu = 0$ and have omitted terms containing
three or more $L$ tensors.  Such terms do in fact not contribute to
the two-pion matrix elements in leading order and next-to-leading
order of ChPT.  To see this, notice that the expansion of $L_\mu$ into
pion fields starts as
\begin{equation}
\label{L-expansion}
L_\mu = i \tau^a \Big( \partial_\mu \pi^a /F 
  - \epsilon^{abc} \pi^b \partial_\mu \pi^c /F^2 
  + \textstyle{\frac{2}{3}} \pi^b\,
       ( \pi^a \partial_\mu \pi^b - \pi^b \partial_\mu \pi^a ) /F^3
  + \mathcal{O}(\pi^3)\, \partial_\mu \pi  \Big) \,.
\end{equation}
Terms with three or more $L$ tensors hence do not contribute to
two-pion matrix elements at Born level.  At one loop, they contribute
only through the tadpole diagram of Fig.~\ref{Fig1}a, where three or
four pion fields from the inserted operator then carry a derivative
contracted with the light-like vector $u$.  The numerator in the
corresponding Feynman integral thus has the form $(uk)^i$, where $i\ge
1$ and $k$ is the loop momentum.  For odd $i$ the integral vanishes
because the integrand is antisymmetric in~$k$, whereas for even $i$
the integral is proportional to $(u^2)^{i/2}$ by Lorentz invariance,
and hence zero as well.
 
Repeating our discussion for the right-handed operators ${\cal
O}^{R}_{\mu_1\ldots\mu_n}$, with 
\begin{equation}
\label{rr}
R_\mu=U \partial_\mu U^\dagger, \qquad\qquad
\chi^R_\pm=\chi U^\dagger \pm U \chi^\dagger
\end{equation}
and definitions analogous to (\ref{L-tensor-def}) and
(\ref{u-notation}), the tensor operator ${\cal O}_n(u)={\cal
O}^{L}_n(u)+{\cal O}^{R}_n(u)$ reads at order $\mathcal{O}(p^0)$ in
ChPT
\begin{equation}\label{On}
{\cal O}_n(u)=c_{n}\, \Big(L_{n}(u)+R_n(u)\Big)
  +\sum_{k=1}^{n-1} c_{n,k}\,
    \Big(L_k(u)\,L_{n-k}(u)+R_k(u)\,R_{n-k}(u)\Big) + \ldots\,,
\end{equation}
where the dots denote terms which do not contribute to two-pion matrix
elements to $\mathcal{O}(p^2)$ accuracy.  The coefficients of the $L$
and $R$ terms must be equal because of parity invariance.  Using the
hermiticity and time reversal properties of ${\cal O}_n(u)$ following
from (\ref{OL-def}), (\ref{OR-def}) and the corresponding properties
of $L_{n}(u)$ and $R_{n}(u)$, one finds
\begin{eqnarray}
\label{cc}
(c_{n})^*=-c_{n} \,, \hspace{0.9em} & & \qquad
\hspace{0.65em} (c_{n})^*=(-1)^n\,c_{n} \,,
\\[2mm]
\label{cc1}
(c_{n,k})^*=c_{n,n-k} \,, & & \qquad
(c_{n,k})^*=(-1)^n\,c_{n,k} \,.
\end{eqnarray}
Neglecting terms which do not contribute to two-pion matrix elements
one derives for the isoscalar and isovector operators at order
${\mathcal O}(p^0)$
\begin{eqnarray}\label{ss}
{\cal O}_n^{s}(u)&=&\sum_{k=1}^{n/2}
a_{n,k}\,\Big(L_k^{a}(u)\,L_{n-k}^{a}(u)
             +R_k^{a}(u)\,R_{n-k}^{a}(u)\Big)\,,
 \ \ \ \ \ \ \ \ \ \ \ \ \ \  (n~\mathrm{even})
\\[1mm]
\label{sv}
 {\cal O}_n^{a}(u)&=&i b_{n,0}\, \Big(L_{n}^{a}(u)+R_n^{a}(u)\Big)
\nonumber\\
&+& \varepsilon^{abc}\sum_{k=1}^{(n-1)/2}
b_{n,k}\,\Big(L_k^{b}(u)\,L_{n-k}^{c}(u)
             +R_k^{b}(u)\,R_{n-k}^{c}(u)\Big)\,,
\ \ \ \ \ (n~\mathrm{odd})
\end{eqnarray}
where $L^a_k(u)$ is defined by $L_k(u)=\tau^a L_k^a(u)$ and $R^a_k(u)$
in analogy.  The expansion coefficients $a_{n,k}$ and $b_{n,k}$ are
linear combinations of the coefficients in (\ref{On}),
\begin{eqnarray}\label{ab}
a_{n,k} &=&c_{n,k}+c_{n,n-k} \hspace{3em} \mbox{for $k\ge 1$} ,
\\[2mm]
b_{n,k} &=&i (c_{n,k}-c_{n,n-k})  
     \hspace{1.85em} \mbox{for $k\ge 1$} , \qquad\qquad
b_{n,0} =-ic_{n \phantom{k}} .
\end{eqnarray}
Because of the symmetry relations (\ref{cc}), (\ref{cc1}) one has
$a_{n,k}=0$ for odd $n$ and $b_{n,k} =0$ for even $n$, and the nonzero
coefficients $a_{n,k}$ and $b_{n,k}$ are real.  To obtain simple
expressions for the form factors $A_{n,k}^I(t)$ in (\ref{GPD-exp}), we
rearrange derivatives and obtain for the isosinglet operator
\begin{equation}\label{ss-n}
{\cal O}_n^{s}(u)=F^2
\sum_{k=0,2,\ldots}^{n-2}\tilde a_{n,k}\, (i u\partial)^k\left(
L^{a}(u)\,(2i u\lrpartial)^{n-k-2}L^{a}(u)+
R^{a}(u)\,(2i u\lrpartial)^{n-k-2}R^{a}(u)
\right)
\end{equation}
with $\lrpartial=\half (\rpartial-\lpartial)$.  Here we used
$L^a_k(u)=(u\partial)^{k-1} L^a_1(u)$ and the abbreviation
$L^a(u)=L^a_1(u)= u^\mu L^a_\mu$ and their analogs for the
right-handed fields.  The coefficients $\tilde{a}_{n,k}$ are linear
combinations of the $a_{n,k}$.  Calculating pion matrix elements one
gets for the isoscalar form factors $A_{n,k}^{I=0}(t)$ at order
${\mathcal O}(p^0)$
\begin{equation}\label{AS}
A_{n,k}^{(0)} =\,
2^{n-k}\Big[\, \tilde a_{n,k-2}-\tilde a_{n,k} \,\Big] ,
\qquad\qquad \mbox{($n$ and $k$ even, $k \le n$)}
\end{equation}
where we set $\tilde{a}_{n,n} =\tilde{a}_{n,-2} =0$.  To ease the
notation we have omitted the isospin label, implying here and in the
following $I=0$ if the first index $n$ of a form factor is even, and
$I=1$ if it is odd.  The inverse of (\ref{AS}) reads
\begin{equation}\label{invAS}
\tilde a_{n,k}=-\sum_{m=0}^{k} 2^{m-n} A_{n,m}^{(0)}
              =\sum_{m=k+2}^{n} 2^{m-n} A_{n,m}^{(0)} \, .
\qquad \mbox{($n$, $m$ and $k$ even, $k \le n-2$)}
\end{equation}
{}From (\ref{AS}) we also see that at order ${\mathcal O}(p^0)$ the
sum of the isoscalar form factors with given $n$ vanishes,
\begin{equation}\label{rest}
\sum_{k=0,2,\ldots}^{n} 2^{k} A_{n,k}^{(0)}=0 \,,
\qquad\qquad \mbox{($n$ even)}
\end{equation}
which according to (\ref{GPD-exp}) gives the soft pion theorem
$H^{I=0}(x,\xi\to 1,0) = 0$ first derived in \cite{Polyakov:1998ze}.
Recalling our discussion at the end of Section~\ref{sec:definitions},
we emphasize at this point that the ChPT equivalent of the twist-two
gluon operators has the same form as (\ref{ss-n}), of course with
different matching coefficients.  Analogs of (\ref{AS}) to
(\ref{rest}) therefore hold for the appropriate moments of the gluon
GPD $H_\pi^g(x,\xi,t)$.

Rewriting (\ref{sv}) in a form similar to (\ref{ss-n}) gives for the
isovector operator
\begin{eqnarray}\label{sv-n}
{\cal O}_n^{a}(u)&=&\tilde b_{n,n-1}\, (i u\partial)^{n-1} V^a(u)
\nonumber\\
&+&
iF^2\epsilon^{abc} \sum_{k=0,2}^{n-3}
   \tilde{b}_{n,k}\, (i u\partial)^k \left(
L^{b}(u)\,(2i u\lrpartial)^{n-k-2}L^{c}(u)+
R^{b}(u)\,(2i u\lrpartial)^{n-k-2}R^{c}(u) \right), \qquad
\end{eqnarray}
where the $\tilde{b}_{n,k}$ are linear combinations of the $b_{n,k}$
in (\ref{sv}).  Here
\begin{equation}
V^a(u) =u^\mu V^a_\mu 
       =-\frac{i}{2} F^2 \Big( L^a(u) + R^a(u) \Big)
\end{equation}
is the isovector vector current already given in (\ref{isov}), which
also implies $\tilde{b}_{1,0} = 1$.  For the isovector form factors
$A^{I=1}_{n,k}(t)$ one derives at $\mathcal{O}(p^0)$
\begin{equation}\label{ASa}
A_{n,k}^{(0)}=
2^{n-k}\Big[\, \tilde b_{n,k}-\tilde b_{n,k-2} \,\Big]\,,
\qquad
\tilde b_{n,k}=\sum_{m=0,2,\ldots}^{k} 2^{m-n} A_{n,m}^{(0)}\,,
\qquad   \mbox{($n$ odd, $k$ even, $k \le n-1$)}
\end{equation}
where we set $\tilde{b}_{n,-2} =0$.  Note that the isovector analog of
(\ref{rest}),
\begin{equation}
  \label{soft-iv}
\sum_{k=0,2,\ldots}^{n-1} 2^{k} A_{n,k}^{(0)}
  = 2^n\, \tilde{b}_{n,n-1} \,,
\qquad\qquad \mbox{($n$ odd)}
\end{equation}
does not have a vanishing right-hand side.  The coefficient $\tilde
b_{n,n-1}$ is however related to another observable, namely the
one-pion matrix element of the isovector axial twist-two operator
\begin{equation}
\mathcal{O}_{5,n}^{a}(u) =
\frac{1}{2} \tr \Big\{ \tau^a \left( 
   \mathcal{O}^R_n(u) - \mathcal{O}^L_n(u) \right) \Big\} =
\frac{1}{2} \bar q\, u^\mu\gamma_\mu\gamma_5\,
   (i u\lrD)^{n-1}\, \tau^{a}\, q \,,
\end{equation}
which to leading order of ChPT has the form
\begin{equation}
  \label{axial-lo}
\mathcal{O}_{5,n}^{a}(u) =\tilde{b}_{n,n-1}
  (i u\partial)^{n-1} A^a(u) +\ldots ,
\end{equation}  
where $A^{a}(u) =u^\mu A_\mu^a=-\frac{1}{2} i F^2\, [ R^{a}(u) -
L^{a}(u)]$ is the isovector axial current and the dots denote terms
which do not contribute to the one-pion matrix element to
$\mathcal{O}(p^2)$ accuracy for reasons analogous to those we
discussed after (\ref{L-expansion}).  One readily finds $\langle
\pi^b(p) |\,\mathcal{O}_{5,n}^{a}(u) \,|0 \rangle = -i\delta^{ab}
(pu)^n F\, \tilde{b}_{n,n-1}$ at order $\mathcal{O}(p^0)$.  The pion
distribution amplitude is defined as
\begin{equation}
\frac{1}{2} \int \frac{d \lambda}{2\pi}\, e^{ix \lambda (p u) /2}\,
  \langle \pi^b(p)|\,  \bar q (-\half\lambda u)\, 
  u^\mu \gamma_\mu\gamma_5\, \tau^a q(\half\lambda u)\, |0 \rangle
 = -i\delta^{ab} F_\pi\, \phi_\pi(x) \,,
\end{equation}
where $\half(1+x)$ is the momentum fraction of the quark in the pion
and $\half \int_{-1}^1 dx\, \phi_\pi(x)=1$.  Here $F_\pi$ is the
physical value of the pion decay constant, given in one-loop ChPT as
\cite{GL}
\begin{equation}
  \label{fpi}
F_{\pi} = F\, 
  \Big( 1 - \frac{m^2}{16\pi^2 F^2}\, \log\frac{m^2}{\mu^2}
          + \frac{m^2}{F^2}\, {l}^r_4(\mu)
          + \mathcal{O}(p^4) \Big) \,,
\end{equation}
where $\mu$ denotes the renormalization scale and $l^r_4(\mu)$ is a
renormalized low-energy constant.  Taking $x$-moments of the
distribution amplitude, one obtains the matrix elements of
$\mathcal{O}_{5,n}^{a}(u)$, so that to order $\mathcal{O}(p^0)$ one
has
\begin{equation}
  \label{da-moments}
B_n \:\stackrel{\mathrm{def}}{=}\: \frac{1}{2^n}
\int_{-1}^1 dx\, x^{n-1} \phi_\pi(x) 
= \tilde{b}_{n,n-1} \,.
\qquad\qquad \mbox{($n$ odd)}
\end{equation}
Combining this with (\ref{soft-iv}) and going from moments to
$x$-space, we obtain the soft pion theorem $H^{I=1}(x,\xi\to 1, 0) =
\phi_\pi(x)$ from \cite{Polyakov:1998ze}.  Let us now consider the
order $\mathcal{O}(p^2)$ corrections to the one-pion matrix elements.
A nonanalytic dependence on the pion mass is generated by one-loop
graphs with the insertion of the leading-order operator
(\ref{axial-lo}), which according to (\ref{L-expansion}) contains
terms with three pion fields.  Since for different $n$ the operators
in (\ref{axial-lo}) only differ by their coefficients
$\smash{\tilde{b}_{n,n-1}}$ and the number of overall derivatives
$u\partial$, the one-loop corrections of the one-pion matrix elements
for all $n$ are universal and hence the same as the corrections to the
isovector axial current, given by (\ref{fpi}).  We can therefore write
\begin{equation}
  \label{axial-nlo}
\langle \pi^b(p)|\, \mathcal{O}_{5,n}^{a}(u) \,|0 \rangle 
  =-i\delta^{ab} (pu)^n F_\pi\, \tilde b_{n,n-1} 
  + \mathcal{O}(m^2)
\end{equation}
to order $\mathcal{O}(p^2)$ accuracy, where the $\mathcal{O}(m^2)$
terms are generated by tree-level insertions of higher-order terms in
the ChPT expansion of $\mathcal{O}^a_{5,n}(u)$, which can be
constructed in analogy to those in Section~\ref{sec:higher-order}.  As
a consequence, the relation (\ref{da-moments}) does not receive
nonanalytic corrections in the pion mass at next-to-leading order in
ChPT, and one has $B_n = \tilde{b}_{n,n-1} + \mathcal{O}(m^2)$ up to
terms of order $\mathcal{O}(p^4)$.  Together with the
$\mathcal{O}(p^2)$ corrections of the form factors $A^{I=1}_{n,k}(t)$,
this will allow us to discuss the one-loop corrections to the soft
pion theorem in Section~\ref{sec:one-loop}.  We note that to the order
given, the bare pion mass $m$ in (\ref{fpi}) and (\ref{axial-nlo}) can
be replaced with its renormalized value $m_\pi$, given by \cite{GL}
\begin{equation}
  \label{mpi}
m_\pi^2 = m^2 
\Big( 1 + \frac{m^2}{32\pi^2 F^2}\, \log\frac{m^2}{\mu^2}
        + \frac{2m^2}{F^2}\, l_3^r(\mu)
        + \mathcal{O}(p^4) \Big) \,.
\end{equation}

Let us finally show that our representations (\ref{ss-n}) and
(\ref{sv-n}) of the operators $\mathcal{O}^s_n(u)$ and
$\mathcal{O}^a_n(u)$ in ChPT are equivalent to those given in
\cite{polyakov1} without derivation.  There, the matching was done for
the nonlocal operators appearing in the definition of GPDs.  For the
left-handed current we find in our notation
\begin{eqnarray}
  \label{kivel-nonloc}
\lefteqn{
\bar{q}_j(-\half\lambda u)\, u^\mu\gamma_\mu \frac{1-\gamma_5}{2}\, 
   q_i(\half\lambda u)
}
\nonumber \\
&\simeq & \frac{1}{2} F^2
  \int_{-1}^1 d\beta \int_{-1+|\beta|}^{1-|\beta|} d\alpha\, 
  F(\beta,\alpha) \,
  U_{ik}^\dagger(\lambda_1 u)\,
  (i u\lrpartial)\, U_{kj}^{\phantom{\dagger}}
  (\lambda_2 u) + \ldots \,,
\end{eqnarray}
where $\lambda_1= -\half (\beta+\alpha) \lambda$ and $\lambda_2 =
\half (\beta-\alpha) \lambda$, and where the dots denote terms not
contributing to two-pion matrix elements at order $\mathcal{O}(p^2)$.
Due to time reversal invariance the function $F(\beta,\alpha)$ is even
in $\alpha$.  Taylor expansion gives the local operators
\begin{eqnarray}
  \label{kivel-loc}
\lefteqn{
\bar{q}_j u^\mu\gamma_\mu \frac{1-\gamma_5}{2}\, 
   (i u\lrD)^{n-1}\, q_i
}
\\
&\simeq & \frac{1}{2} F^2 \sum_{m=0,2,\ldots}^{n-1} 
  \Big( {m \atop n-1} \Big)
  \int_{-1}^1 d\beta \int_{-1+|\beta|}^{1-|\beta|} d\alpha\, 
  \beta^{n-m-1} \alpha^m F(\beta,\alpha) \;
  (\half i u\partial)^{m} \Big[ U_{ik}^\dagger\, 
        (i u\lrpartial)^{n-m} U_{kj}^{\phantom{\dagger}} \Big]
+ \ldots \,,
\nonumber 
\end{eqnarray}
where all fields are taken at space-time position zero and the dots
have the same meaning as before.  By repeated use of the identity
$\lrpartial U = U \lrpartial + U L$ we can rewrite
\begin{equation}
U^\dagger\, (u\lrpartial)^{n}\, U = \left\{
\renewcommand{\arraystretch}{2.4}
\begin{array}{l}
  \displaystyle{\sum_{m=0,2,\ldots}^{n-2}}
  (\half u\partial)^{m} \Big[L\, (u\lrpartial)^{n-m-2}\, L\Big]
  + \mathcal{O}(L^3)
  \qquad\hspace{5.9em} \mbox{($n$ even)} 
\\
  \displaystyle{\sum_{m=0,2,\ldots}^{n-3}}
  (\half u\partial)^{m} \Big[L\, (u\lrpartial)^{n-m-2}\, L\Big]
  + (\half u\partial)^{n-1} L + \mathcal{O}(L^3)
  \qquad \mbox{($n$ odd)} 
\end{array} \right.
\end{equation}
and then project out the isoscalar part for $n$ even and the isovector
part for $n$ odd.  Repeating the same arguments for the right-handed
operators, we recognize precisely the operators that appear in
(\ref{ss-n}) and (\ref{sv-n}).

\section{Tensor operators at order ${\mathcal O}(p^2)$}
\label{sec:higher-order}

The number of independent terms in the ChPT expansion of operators
rapidly grows with the order.  Henceforth we shall consider only the
terms which contribute to two-pion matrix elements at tree-level.
Taking into account the discussion in the previous section, one finds
that the corrections of order ${\mathcal O}(p^2)$ to the operators can
be cast into the form
\begin{eqnarray}\label{DOS}
\Delta{\cal O}_n^{s}(u) \!\! &=& \!\!
\frac{1}{2}\, a_n^{\chi}
\tr\Big\{ \chi_-^{L}\, L_n(u)+\chi_-^{R}\, R_n(u)\Big\} +
\frac{1}{2} \sum_{k=1}^{n/2} a_{n,k}^{\chi}\,
\tr\Big\{L_k(u)\,L_{n-k}(u)+R_k(u)\,R_{n-k}(u)\Big\} \tr\chi_+^L
\nonumber\\[2mm]
&+& \frac{1}{2}\sum_{k=0}^{n/2} \Delta a_{n,k}
\tr\Big\{ L_{\rho,k}(u)\,L_{n-k}^{\rho}(u)
      +R_{\rho,k}(u)\,R_{n-k}^\rho(u)\Big\}
\end{eqnarray}
for ${\cal O}_n^{s}(u)$ and
\begin{eqnarray}\label{DOV}
\Delta{\cal O}_n^{a}(u)&=&
i b_{n,0}^{\chi}\, 
   \Big(L_{n}^{a}(u)+R_n^{a}(u)\Big) \tr\chi_+^L 
 +\varepsilon^{abc}\sum_{k=1}^{(n-1)/2} b_{n,k}^{\chi}\,
\left(L_k^{b}(u)\,L_{n-k}^{c}(u)
     +R_k^{b}(u)\,R_{n-k}^{c}(u)\right) \tr\chi_+^L 
\nonumber\\[2mm]
&+&\varepsilon^{abc}
\sum_{k=0}^{(n-1)/2} \Delta b_{n,k}\,
\left(L_{\rho,k}^{b}(u)\,L_{n-k}^{\rho,c}(u)
     +R_{\rho,k}^{b}(u)\,R_{n-k}^{\rho,c}(u)\right)
\end{eqnarray}
for ${\cal O}_n^{a}(u)$, where we have introduced $L_{\rho,k}(u)=
L_{\rho\mu_1\ldots\mu_k}u^{\mu_1}\ldots u^{\mu_k}$ and its
right-handed counterpart.  Using hermiticity and time reversal
invariance one can deduce that the coefficients $a_n^{\chi}$,
$b_n^{\chi}$, $\smash{a_{n,k}^{\chi}}$, $\smash{b_{n,k}^{\chi}}$ and
$\Delta a_{n,k}$, $\Delta b_{n,k}$ are real.  The order ${\mathcal
O}(p^2)$ part of a twist-two operator which contributes to two-pion
matrix elements is hence parameterized by $2\, ([n/2]+1)$ real
constants.  Note that the number of the form factors $A_{n,k}^I(t)$ in
(\ref{GPD-exp}) is $[n/2]+1$.  At order ${\mathcal O}(p^2)$ the form
factors thus read
\begin{equation}
A_{n,k}^I(t)=A_{n,k}^{(0)}+
A_{n,k}^{(1,m)}\, m^2 + A_{n,k}^{(1,t)}\, t 
+{\mathrm{loop\  contributions}} ,
\end{equation}
where all parameters $A_{n,k}^{(1,m)}$, $A_{n,k}^{(1,t)}$ are
independent and can be expressed in terms of the coefficients
$a_n^{\chi}$, $b_n^{\chi}$ etc.\ in (\ref{DOS}) and (\ref {DOV}).

\section{One-loop contributions}
\label{sec:one-loop}

\begin{figure}
\psfrag{a}[cc][cc]{$a$}\psfrag{b}[cc][cc]{$b$}\psfrag{c}[cc][cc]{$c$}
\centerline{\includegraphics[width=10cm]{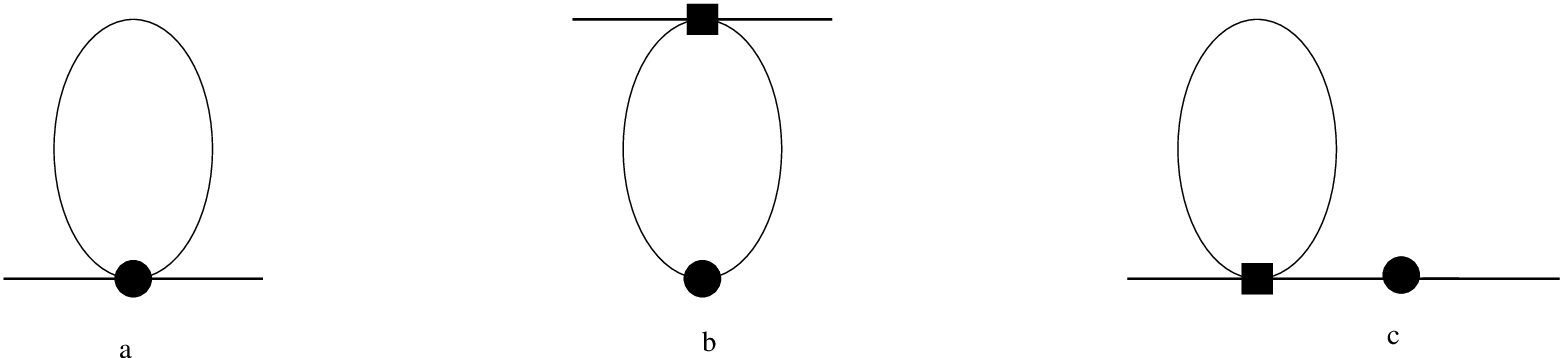}}
\caption{\label{Fig1} The one-loop graphs contributing to the two-pion
  matrix elements of the twist-two operators $\mathcal{O}_s(u)$ and
  $\mathcal{O}_a(u)$. The black blobs denote the operator insertion
  and the black boxes the $\pi^4$ vertex.}
\end{figure}

To calculate the two-pion matrix element of the operator ${\cal
O}_n^{s}(u)+\Delta{\cal O}_n^{s}(u)$ or ${\cal O}_n^{a}(u)+\Delta{\cal
O}_n^{a}(u)$ one has to take into account loop contributions. The
diagrams to be calculated are drawn in Fig.~\ref{Fig1}.  The technique
of such calculations in ChPT is well elaborated, so we omit the
details and directly give our results.  Introducing the notation
$L^a(\lambda u)= u^{\mu} L_\mu^a(\lambda u)$ and its analog for the
right-handed fields one gets
\begin{eqnarray}\label{LL}
\lefteqn{
\langle\pi^{b}(p')|\, L^c(\lambda_1 u)\, L^c(\lambda_2 u)
  \,|\pi^a(p) \rangle^{\mathrm{1~loop}} \;=\;
{}- \delta^{ab}\, \frac{m^2-2t}{32\pi^2 F^4}\, \Gamma(\epsilon)
} \hspace{3em}
\nonumber \\
&& {}\times
\xi^2 (Pu)^2\, e^{i\xi (Pu)\, (\lambda_1+\lambda_2)}
\int_{-1}^1 d\alpha\, e^{i\alpha \xi (Pu)\, (\lambda_1-\lambda_2)}\,
(1-\alpha^2)
\left[\frac{4 m^2-(1-\alpha^2)\, t}{16\pi \mu^2}\right]^{-\epsilon}
\hspace{2em}
\end{eqnarray}
in $d=4-2\epsilon$ dimensions, where $\mu$ is the renormalization
scale.  Taylor expanding in $\lambda_1$ and $\lambda_2$ as we did in
going from (\ref{kivel-nonloc}) to (\ref{kivel-loc}), one can read off
the one-loop corrections for the local operators in (\ref{ss-n}).
Taking into account (\ref{GPD-exp}), one concludes that the form
factors $A_{n,k}^{I=0}(t)$ with $k<n$ do not receive additional
contributions at one-loop order and hence depend linearly on $m^2$ and
$t$,
\begin{equation}\label{A}
A_{n,k}^{I=0}(t)
=A_{n,k}^{(0)}+A_{n,k}^{(1,m)}\, m^2+A_{n,k}^{(1,t)}\, t
\qquad\qquad \mbox{($k\le n-2$)}
\end{equation}
up to corrections of ${\mathcal O}(p^4)$, where as in
Section~\ref{sec:tensor} we omit isospin labels in $A_{n,k}^{(0)}$,
$A_{n,k}^{(1,m)}$ and $A_{n,k}^{(1,t)}$.  For the form factor
$A_{n,n}^{I=0}(t)$ one finds after some algebra
\begin{eqnarray}\label{Ann}
A_{n,n}^{I=0}(t) &=&A_{n,n}^{(0)}+
A_{n,n}^{(1,m)}(\mu)\, m^2+A_{n,n}^{(1,t)}(\mu)\, t
+ \frac{m^2-2t}{64\pi^2 F^2}\, 
\\[1mm]
&\times &\sum_{k=2,4,\ldots}^{n} 2^{k-n} A_{n,k}^{(0)}\,
\int_{-1}^{1} d\alpha\, (1-\alpha^k)\alpha^{n-k}
\left(\log\left[\frac{4m^2-(1-\alpha^2)\,t}{4\mu^2}\right]+1
\right)\,,
\nonumber
\end{eqnarray}
where $A_{n,k}^{(0)}$ is given in (\ref{AS}).  Here
$A_{n,n}^{(1,m)}(\mu)$ and $A_{n,n}^{(1,t)}(\mu)$ are renormalized
using the subtraction scheme of \cite{GL}, where together with a
$1/\epsilon$ pole one subtracts a finite constant in the combination
$1/\epsilon +\psi(2)+\log(4\pi)$.  In particular, we find for $n=2,4$
\begin{eqnarray}
\label{summ}
A_{2,2}^{I=0}(t)&=&A_{2,2}^{(0)}\left(1+\frac{m^2-2t}{48\pi^2F^2}\,
\Big[
\log\frac{m^2}{\mu^2}+\frac{4}{3} -\frac{t+2m^2}{t}\,J(t)
\Big] \right)
\nonumber\\[1mm]
&+& A_{2,2}^{(1,m)}(\mu)\, m^2+A_{2,2}^{(1,t)}(\mu)\, t\,,
\\[2mm]
A_{4,4}^{I=0}(t)&=&A_{4,4}^{(0)}\left(1+\frac{m^2-2t}{40\pi^2F^2}\,
\Big[
 \log\frac{m^2}{\mu^2}+\frac{11}{10} 
+\frac16\sigma^2+\frac{1}{4}(\sigma^4-5)\,J(t)
\Big] \right)
\nonumber\\[1mm]
&+&
A_{4,2}^{(0)}\; \frac{m^2-2t}{960\pi^2F^2}\, \Big[
 \log\frac{m^2}{\mu^2}-\frac{1}{15}
+\sigma^2+\frac{\sigma^2}{2}(3\sigma^2-5)\,J(t)
\Big]
\nonumber\\[4mm]
&+& A_{4,4}^{(1,m)}(\mu)\, m^2+A_{4,4}^{(1,t)}(\mu)\, t\,,
\end{eqnarray}
where
\begin{equation}
  \label{J}
J(t)= 2 + \sigma\log\frac{\sigma-1}{\sigma+1} \,, \qquad\qquad
\sigma=\sqrt{1-\frac{4m^2}{t}}\,.
\end{equation}
For $|t|<4m^2$ one can expand $J(t) = -2/(3\sigma^2) - 2/(5\sigma^4) -
\mathcal{O}(1/\sigma^6) = t/(6m^2) + t^2/(60m^4) -
\mathcal{O}(t^3/m^6)$.  Note that to the order we are working at, one
may replace the bare quantities $F^2$ and $m^2$ in the formulae of the
present section with their one-loop expressions, given in (\ref{fpi})
and (\ref{mpi}).

The order $\mathcal{O}(p^2)$ correction to the matrix element of the
isovector vector current~(\ref{isov}) reads~\cite{GL}~\footnote{The
corresponding form factor is known to two-loop accuracy
now~\cite{BCT}.}
\begin{equation}\label{ISO-one}
\langle\pi^b(p')|\, V_{\mu}^c \,|\pi^a(p) \rangle
=2 i\epsilon^{abc}\, P_\mu 
\left( 1 + F_V(t,m^2) - l_6^r(\mu)\, t /F^2
\right)\,,
\end{equation}
where the renormalized low-energy constant $l_6^r(\mu)$ is defined in
\cite{GL} and
\begin{eqnarray}\label{FV}
F_V(t,m^2) &=&
\frac{1}{128\pi^2F^2} \int_{-1}^1 d\alpha\,
\Big[ 4m^2-(1-\alpha^2)\,t \,\Big]
\log\left[\frac{4m^2-(1-\alpha^2)\,t}{4\mu^2}\right]
-\frac{m^2}{16\pi^2F^2}\log\frac{m^2}{\mu^2}
\nonumber \\[2mm]
&=&
\frac{1}{96\pi^2F^2}\left[(t-4m^2)\,J(t)
 -t\left(\frac{1}{3}  +\log\frac{m^2}{\mu^2}\right)
\right]\,.
\end{eqnarray}
To complete the calculation of the isovector current at order
$\mathcal{O}(p^2)$ it is sufficient to compute the one-loop
corrections to the pion matrix element of the nonlocal operator
${\mathcal O}^{a}(\lambda_1,\lambda_2)=\epsilon^{abc}
L^{b}(\lambda_1u)L^{c}(\lambda_2u)$.  We obtain
\begin{eqnarray}\label{O-ISO}
\lefteqn{
\langle\pi^b(p')|\, {\mathcal O}^{c}(\lambda_1,\lambda_2)
  \,|\pi^a(p) \rangle^{\mathrm{1~loop}}
={}-\frac{m^2}{16\pi^2F^2}\, \Gamma(-1+\epsilon)
\left(\frac{m^2}{4\pi \mu^2}\right)^{-\epsilon}
\langle\pi^b(p')|\, {\mathcal O}^{c}(\lambda_1,\lambda_2)
  \,|\pi^a(p) \rangle^{\mathrm{tree}}
} \hspace{2em}
\nonumber\\[1mm]
&-&
\epsilon^{abc} \frac{1}{64\pi^2 F^4}\,  
\Gamma(-1+\epsilon)\, \xi\, (Pu)^2\, 
e^{i\xi (Pu)\,(\lambda_1+\lambda_2)} 
\nonumber \\[1mm]
&& {}\times
\int_{-1}^1 d\alpha\, \Big[4m^2-(1-\alpha^2)\, t\,\Big]
\left[\frac{4m^2-(1-\alpha^2)\,t}{16\pi \mu^2}\right]^{-\epsilon}
\frac{\partial}{\partial\alpha}
\left[e^{i\alpha \xi (Pu)\, 
     (\lambda_1- \lambda_2)}\,(1-\alpha^2)\right]\,.
\hspace{2em}
\end{eqnarray}
Making use of (\ref{sv-n}) and (\ref{ASa}) one can represent the form
factors $A_{n,k}^{I=1}(t)$ with $k<n-1$ as
\begin{equation}\label{AA}
A_{n,k}^{I=1}(t) = A_{n,k}^{(0)}
\left(1-\frac{m^2}{16\pi^2F^2}\log\frac{m^2}{\mu^2}\right)
+A_{n,k}^{(1,m)}(\mu)\, m^2+A_{n,k}^{(1,t)}\, t
\qquad \mbox{($k \le n-3$)}
\end{equation}
up to corrections of order ${\mathcal O}(p^4)$.  For the form factor
$A_{n,n-1}^{I=1}(t)$ one finds after some algebra
\begin{eqnarray}\label{Annm1}
\lefteqn{
A_{n,n-1}^{I=1}(t)=
A_{n,n-1}^{(0)}
\left(1-\frac{m^2}{16\pi^2F^2}\log\frac{m^2}{\mu^2}\right)
+A_{n,n-1}^{(1,m)}(\mu)\, m^2+A_{n,n-1}^{(1,t)}(\mu)\, t
} 
\\
&+&
\frac{1}{64\pi^2 F^2}\sum_{k=0,2,\ldots}^{n-1} 
2^{k-n}\,(n-k)\, A_{n,k}^{(0)}
\int_{-1}^{1} d\alpha\, \alpha^{n-k-1}
\Big[ 4m^2-(1-\alpha^2)\,t \,\Big]
\log\left[\frac{4m^2-(1-\alpha^2)\,t}{4\mu^2}\right]\,.
\nonumber
\end{eqnarray}
For $n=1$ this gives
\begin{equation}
A_{1,0}^{I=1}(t) = 
2\, \Big( 1+F_V(t,m^2) \Big) + A^{(1,t)}_{1,0}(\mu)\, t \,,
\end{equation}
where $A^{(1,t)}_{1,0}(\mu)= -2l_6^r(\mu) /F^2$ according to
(\ref{ISO-one}).  Note that at $t=0$ this form factor is fixed by
current conservation, so that $A^{(1,m)}_{1,0}(\mu)=0$.  For $n=3$ we
find
\begin{eqnarray}\label{Ison3}
A_{3,2}^{I=1}(t)&=&
A_{3,2}^{(0)}\, \Big(1+F_V(t,m^2)\Big) +
A_{3,2}^{(1,m)}(\mu)\, m^2 + A_{3,2}^{(1,t)}(\mu)\, t
\nonumber\\[2mm]
&-& A_{3,0}^{(0)}\; \frac{t}{1280\pi^2 F^2}\left[
 (5\sigma^2-3)\log\frac{m^2}{\mu^2}+\frac{6}{5}
-\frac{4}3\sigma^2-2\sigma^4 J(t)
\right]\,.
\end{eqnarray}
Our results (\ref{AA}) and (\ref{Annm1}) agree with Eq.~(27) in
\cite{polyakov1}, and our results (\ref{A}) and (\ref{Ann}) agree with
Eq.~(26) there if a misprint is corrected.\footnote{The one-loop term
in Eq.~(26) of \protect\cite{polyakov1} needs to be multiplied with a
factor $1/2$ \protect\cite{Kolya}.}
Furthermore, our expression (\ref{summ}) for $A_{2,2}^{I=0}(t)$ agrees
with \cite{DL91}.

{}From (\ref{A}) and (\ref{Ann}) we can readily generalize to order
$\mathcal{O}(p^2)$ the relation (\ref{rest}) for the sum of isoscalar
form factors of given $n$.  In particular we find for $t=0$
\begin{equation}
  \label{spi-theo-0}
\sum_{k=0,2,\ldots}^{n} 2^{k} A_{n,k}^{I=0}(0) \;=\;
   \frac{m_\pi^2}{64\pi^2 F_\pi^2}\, \log\frac{m_\pi^2}{\mu^2}
   \sum_{k=2,4,\ldots}^{n} 2^{k} A_{n,k}^{I=0}(0)\,
        \frac{2k}{(n+1) (n-k+1)} 
 + \mathcal{O}(m_\pi^2) ,
\end{equation}
where on the right-hand side we have replaced the tree-level
quantities $F$, $m$, $A_{n,k}^{(0)}$ by the full ones.  We have a
relation between observables, which may for instance be tested in
lattice calculations as a consistency test for the applicability of
one-loop ChPT for a given range of unphysical pion masses.  Using
(\ref{AA}) and (\ref{Annm1}) and our discussion following
(\ref{axial-nlo}), we obtain a relation at order $\mathcal{O}(p^2)$
between the sum of isovector form factors of given $n$ and the moments
$B_n$ of the pion distribution amplitude.  For $t=0$ the logarithms in
$m^2$ cancel and we have
\begin{equation}
  \label{spi-theo-1}
\sum_{k=0,2,\ldots}^{n-1} 2^{k} A_{n,k}^{I=1}(0) =
  2^n B_n + \mathcal{O}(m_\pi^2) \,.
\end{equation}
For $n=1$ this relation is trivial since $A_{1,0}^{I=1}(0)=2$ and
$B_1=1$ are exact, but for $n\ge 3$ it presents again an interesting
test in lattice QCD, where both the form factors $A_{n,k}^{I}(t)$ and
the moments $B_n$ are calculable.  Translated to $x$-space, one finds
that the soft-pion theorem $H^{I=1}(x,\xi\to 1,0) = \phi_\pi(x) +
\mathcal{O}(m_\pi^2)$ does not contain nonanalytic terms in the pion
mass at one-loop order, as was pointed out in
\cite{polyakov1}.\footnote{To be precise, in \protect\cite{polyakov1}
this relation was given with the pion distribution amplitude taken in
the chiral limit, without discussing its one-loop properties in ChPT.}

\section{Summary and conclusions}
\label{sec:sum}

We have calculated the moments of the generalized parton distributions
of the pion at order $\mathcal{O}(p^2)$ of ChPT.  Our main results are
given in (\ref{A}), (\ref{Ann}) and (\ref{AA}), (\ref{Annm1}) for the
isoscalar and isovector matrix elements, respectively.  We find that
at one-loop order only the form factors accompanying the highest power
of $\xi$ at given $n$ have a nontrivial dependence on $t$ and
$m_\pi^2$.  All others depend linearly on $t$.  The isoscalar form
factors are linear functions of $m_\pi^2$ as well, whereas the
isovector ones receive universal logarithmic corrections in $m_\pi^2$.
At order $\mathcal{O}(p^2)$ the $[n/2]+1$ form factors $A_{n,k}^{I}$
at given $n$ are determined by $3\, ([n/2]+1)$ real parameters, which
in the isoscalar case are subject to the restriction~(\ref{rest}).
One can check that for a fixed ratio $t/m_\pi^2$ the form factors in
(\ref{Ann}) and (\ref{Annm1}) deviate from a linear behavior in
$m_\pi^2$ only for pion masses close to the physical value, and that
this deviation is rather small.

The possibility to use the formulae derived here to extrapolate
lattice data obtained for pion masses around $500$~MeV to the physical
pion mass obviously depends crucially on the size of the higher-order
corrections.  Taking into account that ChPT works quite well in the
kaon sector~\cite{GL1} one may hope that the two-loop corrections will
not be large for $m_\pi^2\sim 500$~MeV.  In any case a strong
deviation of the lattice data from a linear dependence on $m_\pi^2$
would indicate the need for higher-order ChPT fits to describe the
lattice data.  The comparison of the ChPT predictions with lattice
data for the form factor $A_{1,0}^{I=1}(t) = 2
F_\pi^{\mathit{em}}(t)$, where all $\mathcal{O}(p^2)$ parameters are
known~\cite{GL,BCT}, will provide a crucial convergence test.  The
same holds for relations which involve only lattice observables,
namely the relation (\ref{spi-theo-0}) between isoscalar form factors
and the relation (\ref{spi-theo-1}) between isovector form factors and
the moments of the pion distribution amplitude, as well as their
generalizations for finite $t$.

At two-loop order six new parameters appear for each form factor.
Three of them are the coefficients of the analytic terms $m_\pi^4$, $t
m_\pi^2$, $t^2$, and the other three arise from terms in the expansion
of operators which do not contribute to the two-pion matrix element at
one-loop order ($m_\pi^4 \log^2 m_\pi^2$, $m_\pi^4 \log m_\pi^2$, and
$t m_\pi^2\ \log m_\pi^2$).  Obviously the lattice data, even if it
covers several different volumes and many pion masses, can hardly be
hoped to be accurate enough to obtain a stable simultaneous fit of all
these parameters.  However, the same parameters should enter the
$m_{\pi}$ and volume dependence of other lattice observables, such
that these will provide complementary information.  A detailed
analysis of these relationships remains, however, to be worked out.

\begin{center}
{\bf Acknowledgments}
\end{center}

We are indebted to D. Br\"ommel, N.~Kivel, D. M\"uller, and especially
to Th.~Hemmert for very helpful discussions.  This work was supported
by the Helmholtz Association, contract number VH-NG-004 and by the RFFI
grants 03-01-00837 (A.~M.).


\end{document}